\documentstyle[12pt]{article}

\def\be{\begin{equation}}
\def\ee{\end{equation}}
\def\ben{\begin{displaymath}}
\def\een{\end{displaymath}}
\def\ba{\begin{array}{c}}
\def\bal{\begin{array}{l}}
\def\ea{\end{array}}

\begin{document}

\titlepage
\vspace*{2cm}

 \begin{center}{\Large \bf
 Coupled-channel version of

 PT-symmetric square well
  }\end{center}

\vspace{10mm}

 \begin{center}
Miloslav Znojil

 \vspace{3mm}

\'{U}stav jadern\'e fyziky AV \v{C}R, 250 68 \v{R}e\v{z}, Czech
Republic\footnote{e-mail: znojil@ujf.cas.cz}

\end{center}

\vspace{5mm}


\section*{Abstract}

Coupled pair of PT-symmetric square wells is studied as a
prototype of a quantum system characterized by two manifestly
non-Hermitian commuting observables.  Via the diagonalization of
our Hamiltonian $H\neq H^\dagger$ and spin-like observable $\Omega
\neq \Omega^\dagger$ we demonstrate that there exists a domain of
couplings where both the respective sets of eigenvalues $E_n$, $n
= 0, 1, \ldots$ (energies) and $\sigma=\pm 1$ (``spin
projections") remain {\em real}. In such a ``measurable" regime
the model acquires a consistent probabilistic interpretation
mediated by our selection of one of many available
interaction-dependent scalar products.

\vspace{5mm}

PACS

03.65.Ca
03.65.Ge;

\newpage

\section{Introduction}

One of the keys to the proposal of PT-symmetric Quantum Mechanics
(PTSQM) by Bender and Boettcher \cite{BB} lied in the reality of
the spectrum of the imaginary one-dimensional oscillator well
$V_{(cubic)}(x) \sim ix^3$. Although the rigorous confirmation of
that fundamental as well as phenomenologically welcome property of
the model has been delivered a few years later \cite{DDT}, the
proof remains rather abstract and complicated \cite{Shin}. For
this reason, a lot of parallel attention has been paid to the
other, exactly solvable non-Hermitian potentials with real spectra
\cite{sgezou}. People studied partially solvable (often called
quasi-exact) analytic alternatives to $V_{(cubic)}(x) \sim ix^3$
\cite{QES} as well as non-analytic square-well potentials of
similar type \cite{sqw} and their singular point-interaction
limits \cite{Kurasov}.

Solvable choices proved particularly suitable for illustrative
purposes. Their study clarified that PTSQM formalism may be
understood as a very natural extension of Quantum Mechanics, not
asking for any new formulation of the ``first principles". For a
review we may recommend the recent dedicated Workshops'
proceedings \cite{reviews}.

We intend to broaden the scope of the current PT-symmetric models
beyond their popular ordinary differential equation (ODE)
framework. We feel motivated by the observation that the majority
of existing applications of the innovative PTSQM formalism
concerns systems characterized by a {\em single} physical
observable. We intend to fill the gap by an introduction of a
model possessing a {doublet} of commuting independent observables
(sect. 2). Our square-well-type model is solvable and intuitively
transparent (sect. 3). It exemplifies a number of generic features
of PTSQM systems (cf. discussion in sect. 4). Its appeal and
properties are summarized in sect.~5.

A number of technical details is separated in Appendices A (an
account of the perturbation representation of energies), B
(summarizing the PTSQM formulism in a modified Dirac's notation),
C (on norms) and D (on the single-channel projection).

\section{The model}

\subsection{${\cal PT}-$symmetry and its generalizations}

The productivity of the counterintuitive PTSQM approach has been
mainly revealed via studies of specific, concrete examples. Many
of them proved too exceptional. Typically, one may recollect the
elementary, exactly solvable spiked harmonic oscillator of
ref.~\cite{ptho} with its complete confluence of all the
infinitely many 'exceptional points' defined as the couplings at
which two neighboring real energies merge and complexify
\cite{Heiss}.

Among the less exceptional toy models, many conjectures have been
deduced from discontinuous solvable potentials. The simplest,
purely imaginary piece-wise constant potential of ref.~\cite{sqw}
with single discontinuity contributed to our understanding of the
mechanisms of stabilization of the real spectra \cite{frag}.
Supersymmetric partners of this potential have been found
obtainable by non-numerical means \cite{Bijan}. The study of its
physical aspects and classical limit proved facilitated by its
perturbative tractability \cite{Batal}. A model-independence of
most of these observations was confirmed by the long-range
square-well model with two discontinuities \cite{robust}, by the
short-range model with three discontinuities \cite{Bila} and by
the harmonic oscillator decorated with two delta-function
discontinuities \cite{Demiralp}.

All the above-mentioned solvable models offered an independent
support for the inspiring conjecture of ref.~\cite{BB} that the
reality of spectra is related to the so called PT-symmetry of the
potentials. This connects the observed absence of the complex
energy eigenvalues with the invariance of the Hamiltonians with
respect to the combined action of a complex conjugation ${\cal T}$
and a parity reversal ${\cal P}$, ${\cal PT}H = H{\cal PT}$ (cf.
ref.~\cite{BG}).

The latter type of symmetry gave its name to all the PTSQM
formalism. Beyond the simplest ODE Hamiltonians, the emphasis on
the parity and time reversal meaning of the operator ${\cal PT}$
may be weakened \cite{BBJ}. Thus, the parity ${\cal P}$ may be
replaced by an arbitrary invertible self-adjoint operator $P$ or,
in a less confusing notation, $\theta$. In parallel, the meaning
of the complex conjugation ${\cal T}={\cal T}^{-1}$ may be
extended to all the Hermitian-conjugation involutions $A \to
A^\dagger={\cal T} A {\cal T}^{-1}$ \cite{Critique}. The PTSQM
concepts become applicable to nonsymmetric operators and the
PT-symmetry becomes re-interpreted as the property
 \be
 H^\dagger = \theta\,H\,\theta^{-1}, \ \ \ \ \ \ \
 \theta=\theta^\dagger
 \label{thetapseudohermitian}
 \label{rovdva}
 \ee
called $\theta-$pseudo-Hermiticity of $H$ \cite{ali}.

An appreciation of the subtlety of the latter generalization
requires a non-ODE model possessing more than one observable. We
intend to describe here such a model in detail.

\subsection{Two coupled channels}


The frequent use of the coupling of channels in physics
\cite{KG,KGA} attracted our attention to the partitioned
 \be
 \theta =\theta^\dagger=\left (
 \begin{array}{cc}
 0&{\cal G}\\
 {\cal G}&0
 \ea
 \right ), \ \ \ \ \
 \theta^{-1}=\left (
 \begin{array}{cc}
 0&{\cal G}^{-1}\\
 {\cal G}^{-1}&0
 \ea
 \right )
 \label{provazanekanaly}
 \ee
with any parity-type invertible sub-operator ${\cal G}={\cal
G}^\dagger$ which is not necessarily involutive.

A coupled pair of equal-mass particles moving in single spatial
dimension inside a deep square-well box will be considered,
exhibiting the symmetry (\ref{rovdva}) + (\ref{provazanekanaly})
of their non-Hermitian Hamiltonian $H = H_{(kinetic)} +
H_{(interaction)}$. In units $\hbar = 2m = 1$ we shall have
 \be
 H_{(kinetic)} =
 \left (
 \begin{array}{cc}
 -\frac{d^2}{dx^2}&0\\
0&
 -\frac{d^2}{dx^2}
 \ea
 \right ),
 \ \ \ {\rm} \ \ \
 H_{(interaction)}=
 \left (
 \begin{array}{cc}
 V_a(x)&W_b(x)\\
 W_a(x)&V_b(x)
 \ea
 \right )\,.
 \label{vazanekanaly}
 \ee
The pseudo-metric (\ref{provazanekanaly}) commutes with the
kinetic (i.e., differential) operator $H_{(kinetic)}$ so that the
$\theta-$pseudo-Hermiticity condition~(\ref{thetapseudohermitian})
will degenerate to an explicit definition of $V_b={\cal
G}^{-1}V_a^\dagger {\cal G}$ and to the two ${\cal
G}-$pseudo-Hermiticity relations
 \ben
 W^\dagger_a={\cal G}W_a{\cal G}^{-1}
 , \ \ \ \ \ \ \ \
 W^\dagger_b={\cal G}W_b{\cal G}^{-1}.
 \een
Although we emphasized the generality of the operator ${\cal G}$
acting in the single-channel subspace, we shall simplify the
discussion by a return to the common {parity reversal} in what
follows, ${\cal G}={\cal P}$.

In order to select a specific channel-coupling interaction in
(\ref{vazanekanaly}) we shall pick up a maximally simplified,
purely imaginary piece-wise-constant potential such that ${\rm
Re}\,V_{a,b}(x) ={\rm Re}\,W_{a,b}(x) =0$ and
  \be
 \bal
 {\rm Im}\,W_a(x) =X >0, \ \ \ \ \ {\rm Im}\,W_b(x) =Y>0, \ \ \ \ \
  x \in (-1,0),\\
 {\rm Im}\,W_a(x) =-X, \ \ \ \ \ {\rm Im}\,W_b(x) =-Y, \ \ \ \ \ x \in
 (0,1),\\
 {\rm Im}\,V_a(x) = {\rm Im}\,V_b(x) =Z, \ \ \ \ \
  x \in (-1,0),\\
 {\rm Im}\,V_a(x) = {\rm Im}\,V_b(x) =-Z, \ \ \ \ \
  x \in (0,1).
 \ea
 \label{SQW}
 \ee
This non-Hermitian model of the coupling of channels is defined in
terms of its three real parameters $X$, $Y$ and $Z$. This
represents an immediate generalization of the single-channel
square well of ref.~\cite{sqw} in which the spectrum happened to
be real at all the not too large coupling constants \cite{sqwge}.
Basically, we intend to prove the same for eq.~(\ref{SQW}).

\section{Solutions}

\subsection{Non-Hermitian symmetry $\Omega$}

Hamiltonian $H$ of eq.~(\ref{vazanekanaly}) enters the
coupled-channel Schr\"{o}dinger equation
 \be
 H\,
 \left (
 \ba
 \varphi(x)\\
 \chi(x)
 \ea
 \right )
 =
 E\,\left (
 \ba
 \varphi(x)\\
 \chi(x)
 \ea
 \right ).
 \label{SE}
 \ee
It specifies the bound states of the model when accompanied by the
current asymptotic  boundary condition re-scaled to $L = 1$,
 \be
 \left .
 \left (
 \ba
 \varphi(x)\\
 \chi(x)
 \ea
 \right )
 \right |_{x=\pm L}
 =\left (
 \ba
 0\\0
 \ea
 \right ).
 \label{bc}
 \ee
A  merit of such a choice of the example is that its two-by-two
Hamiltonian of eq.~(\ref{vazanekanaly}) commutes with the
spin-like constant matrix
 \ben
 \Omega=
 \left (
 \begin{array}{cc}
 0& \omega^{-1}\\
 \omega&0
 \ea
 \right ) , \ \ \ \ \ \ \ \omega=\sqrt{\frac{X}{Y}}>0.
 \een
It plays the role of another non-Hermitian ``observable". Its
eigenvalues $\sigma = \pm 1$ are real and it exhibits also the
$\theta-$pseudo-Hermiticity property,
 \ben
 \Omega^\dagger =
 \theta \Omega\theta^{-1}.
 \een
The existence of the symmetry $\Omega$ implies that
Schr\"{o}dinger equation (\ref{SE}) may be complemented by the
fixed-spin constraint
 \be
 \Omega\,
 \left (
 \ba
 \varphi_{\sigma}(x)\\
 \chi_{\sigma}(x)
 \ea
 \right )=
 \sigma\,
 \left (
 \ba
 \varphi_{\sigma}(x)\\
 \chi_{\sigma}(x)
 \ea
 \right ).
 \label{spinarch}
 \ee
It gives the relation between the channels at both the spin
eigenvalues $\sigma = \pm 1$,
 \be
 \chi_{\sigma}(x) = \sigma \omega \varphi_{\sigma}(x).
 \label{bnemod}
 \ee
For the sake of brevity we shall mostly drop the subscripts
$_\sigma$ in what follows.

\subsection{Wavefunctions \label{3.2} }

The connection (\ref{bnemod}) between the channels reduces the
system of equations (\ref{SE}) into the {\em single},
$\sigma-$dependent linear differential equation with the
piece-wise constant coefficients,
 \be
 -\frac{d^2}{dx^2}\varphi_{n}(x)
 +\left [ V_a(x)
 +\sigma\omega W_{b}(x) \right ] \varphi_{n}(x)=
 E_{n}\varphi_{n}(x), \ \ \ \ \ x \in (-1,1).
 \label{SEd}
 \ee
The necessary incorporation of the ``asymptotic" boundary
conditions (\ref{bc}) reduces further its general solutions to the
ansatz
 \be
 \ba
 \varphi(x)= \left \{
 \begin{array}{ll}
  A\,\sin \kappa_L(x+1)
 , \ \ & x \in (-1,0), \\
 C\,\sin \kappa_R(1-x),
  \ \ & x \in (0,1).
 \ea
 \right .
 \ea
 \label{ansatzf}
 \ee
The insertion of this ansatz in eq.~(\ref{SEd}) gives the linear
relations
 \be
 E=\kappa_L^2+{\rm i}(Z+\sigma \sqrt{X Y})
 =\kappa_R^2-{\rm i}(Z+\sigma \sqrt{X Y})
 \label{relaj}
 \ee
which define the energy and connect $\kappa_L$ with $\kappa_R$ at
any given, fixed value of $\sigma = \pm 1$.

Once we expect that the energies are observable we have to {\em
assume} that all their values $E=E_n$ remain real. {\it Vice
versa}, we are persuaded that for a certain fairly broad class of
the coupled-channel non-Hermitian interactions and equations on a
finite interval the {\em general} rigorous proof of the existence
of a non-empty physical domain of parameters (where the energies
remain real) may be based on the straightforward extension of the
proof delivered by Langer and Tretter in the single-channel case
\cite{Langer}. For our present purposes we shall feel satisfied by
the less ambitious approach paralleling simply the single-channel
construction of ref.~\cite{sqw}.

Within our physical domain ${\cal D}$ of the real $X$, $Y$, $Z$
and $E$ the inspection of eqs.~(\ref{relaj}) and (\ref{ansatzf})
reveals that we may set $\kappa_L =\kappa_R^* =\kappa =s-{\rm
i}t$. At a fixed spin $\sigma$ this re-defines
 \be
 E = s^2 - t^2, \ \ \ \ \ \  Z=2st-\sigma\sqrt{XY}
 \label{ansatzfg}
 \ee
in terms of some new pair of real parameters $s$ and $t$. We may
keep one of them positive (say, $s > 0$) while the second one lies
on a branch of a hyperbolic curve,
 \be
  t=t_\sigma (s) = \frac{1}{2s}\,Z_{ef\!f}(\sigma), \ \ \
  \ \ \
Z_{ef\!f}(\sigma) = Z+\sigma\sqrt{XY}, \ \ \
  \sigma = \pm 1.
 \label{rooty}
 \ee
We see that the energies remain non-degenerate with respect to the
spin $\sigma$ in general.

The quantization will be mediated by the requirement of the
continuity of the wave functions $\varphi(x)$ and $\chi(x)$ and of
their first derivatives at $x=0$. These conditions degenerate to
the single pair of complex equations
 \ben
 A \sin \kappa =
 C \sin \kappa^*,\ \ \ \ \ \ \ \ \
 A \kappa \cos \kappa =
 -C \kappa^* \cos \kappa^*.
 \een
The first item fixes the normalization ($A=C\sin \kappa^* / \sin
\kappa$) while the elimination of $C$ gives the complex constraint
${\rm Re}\,(\kappa^{-1}\tan \kappa)=0$. It is equivalent to the
single, $\sigma-$independent real secular equation
 \be
 s\,\sin 2s + t\,\sinh 2t = 0.
 \label{secular}
 \ee
Our construction of bound states is completed. They are determined
by formulae (\ref{ansatzf}) and (\ref{ansatzfg}) while their free
parameters $s$ and $t$ must be fixed by the pair of eqs.
(\ref{rooty}) and (\ref{secular}). A few comments on the practical
numerical and perturbative evaluation of the roots $(s_n, t_n)$
may be found in Appendix A.

\section{Interpretation of the solutions}

One does not leave the Standard Textbook Quantum Mechanics (STQM)
whenever feeling satisfied by the Hilbert space ${\cal
H}_{(physical)}$ where the scalar product [i.e., metric operator
$\Theta$ in its definition $(a,b)_{(physical)}= \langle
a|\Theta|b\rangle$] is kept trivial, $\Theta_{(STQM)} \equiv I$.
In contrast, one is allowed and advised to admit a nontrivial
metric within PTSQM framework, $\Theta_{(PTSQM)} \neq I$
\cite{ook}.

In the latter setting it is important to keep in mind that the
non-Hermiticity of the operators of observables (i.e., of $H$ and
$\Omega$ in our present illustrative example) might lead to some
confusion in the standard Dirac's `bra-ket' notation. For this
reason, the slightly modified `brabra-ket' notation of
ref.~\cite{KGA} is advocated and summarized here in Appendix B.

\subsection{Physical metric $\Theta$ \label{5.1}}

The extended flexibility of PTSQM formalism is compensated by the
necessity of an explicit construction of a consistent physical
metric operator $\Theta_{(PTSQM)}\neq \theta$. It must be
Hermitian ($\Theta=\Theta^\dagger$) and positive definite
($\Theta> 0$). The new freedom broadens the class of the
observables $H={\cal O}_{(PTSQM)}^{(j)}$, $j = 1, 2, \ldots, M$
which must be quasi-Hermitian, i.e., by definition, Hermitian in
our new metric,
 \be
 H^\dagger = \Theta\,H\,\Theta^{-1}.
 \label{quasihermitian}
 \label{rovone}
 \ee
According to the review paper \cite{Geyer} the introduction of the
observables of this type may be made in a mathematically
consistent as well as phenomenologically appealing manner. From a
pragmatic point of view, it proved particularly productive in
nuclear physics where $M> 1$ as a rule.

Requirement (\ref{rovone}) looks difficult to satisfy, especially
in less elementary quantum systems. The key to the {\em technical
feasibility} of the transition STQM $\to$ PTSQM has been found in
the existence of an indeterminate, {\em auxiliary} pseudo-metric
$P$ (re-named as $\theta$, in our coupled-channel
eq.~(\ref{rovdva}), in an attempt to avoid its easy confusion with
a very similar symbol ${\cal P}$ for parity). This means that one
should speak, strictly speaking, about a $\theta{\cal
T}-$symmetric Quantum Mechanics throughout our text.

One usually proceeds in an opposite direction, from a (preferably,
{\em very} simple) pseudo-metric to metric. Thus, a complete set
of eigenstates of a given set of $\theta-$pseudo-Hermitian
observables is constructed in the first step, and the proof of the
reality of the energies is then added as quite a difficult task.
In this sense, our present solvable model may serve as a source of
an insight in the properties of the wavefunctions (cf. sect.
\ref{3.2}) as well as of the domain of the reality of the spectrum
(cf. Appendix A).

On this background let us now address the correct physical
interpretation of the theory. It is to be achieved via a
specification of {the physical metric}, {\em to be constructed} as
a Hermitian and positive definite {\em solution} $\Theta =
\Theta^\dagger > 0$ of eq.~(\ref{quasihermitian}). Its knowledge
will enable us to treat any {quasi-Hermitian} operator $A$ with
the property $A^\dagger=\Theta\,A\,\Theta^{-1}$ as an observable.

Once we start from the spectral representation (\ref{spectral}) of
$H$ (cf. Appendix B), we may recall the property $H^\dagger \Theta
= \Theta H$ and infer that
 \be
 \Theta = \sum_{E\!,\,\sigma\!,\,F\!,\,\tau}\
 |F,\tau\rangle\!\rangle\ S_{F\!,\tau\!,\,E\!,\,\sigma}\
 \langle\!\langle E,\sigma| \,.
 \label{onestare}
 \ee
The choice of the expansion coefficients $S$ must remain
compatible with eq.~(\ref{quasihermitian}) and with the symmetry
$\Omega$. This gives the conditions
 \ben
  S_{E\!,\sigma\!,\,F\!,\,\tau}\
 \left (E^*-F\right )=0, \ \ \ \ \
  S_{E\!,\sigma\!,\,F\!,\,\tau}\
 \left (\sigma-\tau \right )=0.
 \een
The spectrum of energies is assumed real so that the off-diagonal
part of the array $S$ must vanish. We may replace the quadruple
sum (\ref{onestare}) by the double-sum ansatz
 \be
 \Theta = \sum_{n\!,\,\sigma}\
 |E_n,\sigma\rangle\!\rangle\ S_{n\!,\,\sigma}\
 \langle\!\langle E_n,\sigma| \, .
 \label{capitaltheta}
 \ee
It represents the formal solution of eq.~(\ref{quasihermitian})
and contains the infinite sequence of arbitrary coefficients
$S_{n\!,\,\sigma}$. Whenever they do not vanish,
$S_{n\!,\,\sigma}\neq 0$, the operator (\ref{capitaltheta}) is
formally invertible,
 \ben
 \Theta^{-1} = \sum_{n\!,\,\sigma}\
 |E_n,\sigma\rangle\
 \frac{1/S_{n\!,\,\sigma}}{
 \langle E_n,\sigma|E_n,\sigma\rangle\!\rangle\cdot
  \langle\!\langle E_n,\sigma|E_n,\sigma\rangle
 }\
 \langle E,\sigma| \ .
 \een
The necessary \cite{Geyer} Hermiticity of $ \Theta$ is guaranteed
when all the parameters $S_{n\!,\,\sigma}$ remain real. The
necessary positivity of $\Theta$ (which means its tractability as
a genuine physical metric) will be achieved whenever all the
coefficients remain positive,  $S_{n\!,\,\sigma}>0$.

We see that the acceptable metric (\ref{capitaltheta}) compatible
with all the standard physical requirements exists (at least in
the formal sense) and is non-unique. One is allowed to impose some
other mathematical or physical requirements \cite{BBJ,Geyeru}.

\subsection{Normalization conventions and the norm
\label{5.1bb}}

In our bound-state solutions $|E_n,\sigma\rangle $ of
eq.~(\ref{ansatzf}) we are free to use any complex `normalization'
constants $C=C_{n,\sigma}$. The same freedom of choice applies to
another series of the `normalization' constants which would appear
in the similar formulae for the `left' eigenkets $\langle\!\langle
E_n,\sigma|$. This is unaffected by the observation of Appendix B
that the respective definitions (\ref{u}) and (\ref{v}) are
connected by the pseudo-Hermiticity property
(\ref{thetapseudohermitian}). One can only conclude that in all
the non-degenerate cases the following proportionality rule
remains valid,
 \ben
  |E,\sigma\rangle \sim \theta^{-1} \,|E^*,
 \sigma^* \rangle \! \rangle.
 \een
For $E=E^*$ and $\sigma=\sigma^*$ at $Z < Z_{crit}$ we may treat
the latter rule as a {\em definition} of the eigenkets
$|\cdot,\cdot \rangle\!\rangle \in {\cal H}$ up to a
normalization,
 \be
 |E,\sigma\rangle\!\rangle =\theta\,
 |E,\sigma\rangle\,{\varrho}_{E\sigma}^{(optional)}\,, \ \ \ \ \ E=E_0, E_1, \ldots =
 {\rm real},
 \ \ \ \ \sigma = \pm 1.
 \label{star}
 \ee
The explicit solution of the left eigenproblem  (\ref{v}) is made
redundant but the freedom in the choice of a convenient relative
normalization (RN) factor ${\varrho}_{E\sigma}^{(optional)}$
survives.

The same factor  emerges in the following formula for the overlaps
between eigenstates,
 \be
 \langle\!\langle E_{n'},\sigma'|E_n,\sigma\rangle=
 \delta_{\sigma\sigma'}
 \delta_{n n'}\,{\varrho}_{E_n \sigma}^{(optional)}\,
 \langle E_n,\sigma|\,\theta\,|E_n,\sigma\rangle\,,
 \ \ \ \  n,n'=0,1, \ldots,
 \ \  \ \sigma, \sigma' = \pm 1.
 \label{capibara}
 \ee
Before its deeper analysis one may consult Appendix C which shows
how our coupled-channel Hilbert space ${\cal H}$ may be
partitioned into two single-channel subspaces ${\cal H}_c$. This
partitioning is prescribed there in such a way that our (free)
choice of an overall normalization coefficient entering the right
eigenstate $|E_n,\sigma\rangle$ is unambiguously inherited by its
single-channel components $ |\varphi_n\rangle$ via
eq.~(\ref{bcmod}). In parallel, an {\em independent} choice of the
coefficient in each left eigenstate $\langle\langle E_n,\sigma|$
is transferred to its single-channel components by
eq.~(\ref{leftpar}).

On this background we must inter-relate our full-space and
subspace RN conventions. The factor ${\varrho}_{E_n
\sigma}^{(optional)}$ introduced in the full space ${\cal H}$ may
differ from its subspace partner ${\lambda}_{E_n
\sigma}^{(optional)}$ of eq.~(\ref{starlet}). By construction,
fortunately, both these quantities happen to coincide [cf.
eq.~(\ref{coincide}) and the rest of Appendix C for more details
of the proof]. As a consequence, the partitioning enables us to
replace the non-vanishing overlaps in (\ref{capibara}) by the much
simpler matrix elements (\ref{subarujika}).

An inspection of the formulae (\ref{capibara}) and
(\ref{subarujika}) reveals that the absolute value of the
self-overlap $\langle\!\langle E,\sigma|E,\sigma\rangle$ may be
re-scaled to one by an appropriate choice of the `normalization'
constants $C=C_{n,\sigma}$ in eq.~(\ref{ansatzf}). We also
restrict the RN factors by the similar condition while their sign
remains free, ${\varrho}^{(optional)}_{E_n\sigma}=\pm 1$. The key
consequence lies in the fact that the sign of all the
non-vanishing self-overlaps (\ref{subarujika}) is {\em fully}
controlled by the sign of our optional RN factor. In an opposite
direction, we have a  {\em freedom to prescribe} such a specific
set ${\varrho}^{(special)}_{E_n\sigma}=\pm 1$ which {guarantees}
the {\em positivity} of all the `special' self-overlaps. Thus, in
the light of eq.~(\ref{subarujika}) we simply postulate
$\langle\!\langle E_n,\sigma|E_n,\sigma\rangle_{(special)} > 0$,
i.e.,
  \be
   \sigma\,{\varrho}_{E_n,\sigma}^{(special)}\,\langle n |{\cal P}|  n
  \rangle>0\,.
  \label{bezejm}
 \ee
After one evaluates the matrix element, this equation defines the
{\em dynamically determined} ``physical" RN factors which make our
basis biorthonormal,
 \ben
 \langle\!\langle E_{n'},\sigma'|E_n,\sigma\rangle_{(special)}=
 \delta_{\sigma\sigma'}
 \delta_{nn'}
 \,.
 \een
The (positive-definite) ``physical" norm
 \ben
 |\!|\,|E_n,\sigma\rangle |\!|_{(physical)}=
 \sqrt{ \langle\!\langle E,\sigma|E,\sigma\rangle_{(special)}}\,
 \een
of our bound states is obtained as a byproduct. This definition
may be extended to all the elements of ${\cal H}$ via completeness
relations (\ref{completeness}) \cite{BBJ,ali,psunit}.

\subsection{Quasi-parity ${\cal Q}$
\label{5.1b}}

In the literature people call the re-scaled RN coefficients a
``charge" \cite{BBJ} or ``quasi-parity" \cite{ja}. It is important
to notice that in our present example their explicit determination
is not difficult since the matrix elements $\langle n |{\cal P}| n
\rangle$ may be evaluated in closed form. Due to the purely
trigonometric character of the wavefunctions (\ref{ansatzf}) we
have
 \ben
  \frac{1}{AA^*}\,\langle n |{\cal P}|  n \rangle
  =\int_{-1}^0
  \sin \kappa^*(x+1)\sin \kappa(-x+1)\,dx +
  \int_{0}^1
  \sin \kappa(-x+1)\sin \kappa^*(x+1)\,dx=
  \een
  \ben
  =\frac{1}{2s} \sin 2s\, {\rm cosh} 2t -
  \frac{1}{2t} \cos 2s\, {\rm sinh} 2t
  \,.
 \een
The $n-$dependence of this element is particularly transparent at
the higher excitations with large $s= {\cal O}(n)$ and small $|t|=
{\cal O}(1/n)$. Our exact formula degenerates to its leading-order
estimate
 \ben
  \frac{1}{AA^*}\,\langle n |{\cal P}|  n \rangle
 =
  - \cos 2s+
  {\cal O}\left ( \frac{1}{n} \right )
 = (-1)^n
   \cos Q_n+
  {\cal O}\left ( \frac{1}{n} \right )
 = (-1)^n
  +
  {\cal O}\left ( \frac{1}{n} \right )
  \,.
 \een
Thus, the ``charge" or ``quasi-parity"  is specified by the closed
formula
 \ben
 {\varrho}^{(special)}_{E_n \sigma}=
  (-1)^n
 \sigma, \ \ \ \ \ \ \ \sigma = \pm 1, \ \
 \ \ \ \ \ \ \ n \gg 1
 \een
at the higher excitations.

Quasi-parities may now be interpreted as eigenvalues of a certain
operator ${\cal Q}$,
 \be
 {\cal Q}
 |E_n,\sigma\rangle=
 |E_n,\sigma\rangle\,
 {\varrho}^{(special)}_{E_n \sigma}
 \,.
  \label{defina}
 \ee
We insert eq.~(\ref{defina}) in (\ref{star}) and deduce that
 \be
 \langle\!\langle E_{n'},\sigma'|E_n,\sigma\rangle_{(special)}=
 \langle E_{n'},\sigma'|\theta\,{\cal Q}  |E_n,\sigma\rangle,
 \ \ \ \  n,n'=0,1, \ldots,
 \ \  \ \sigma, \sigma' = \pm 1.
 \label{expression}
 \ee
This identification defines an overlap of {\em two different}
vectors using a specific scalar product in ${\cal H}$. In the
light of sect. \ref{5.1} such a particular product corresponds to
a particular metric operator,
 \ben
\Theta_{(particular)}=
 \theta\,{\cal Q}.
 \een
Combining this relation with eq.~(\ref{capitaltheta}) we get
 \be
 S_{n\!,\,\sigma}^{(special)}=\frac{1}{
 \langle\!\langle E_n,\sigma|E_n,\sigma\rangle_{(special)}} \, .
 \label{ucapaltheta}
 \ee
{\it Vice versa}, the violation of the one-to-one correspondence
(\ref{ucapaltheta}) between the metric and norm would require an
``artificial" introduction of an $n-$ and $\sigma-$dependence into
our definition (\ref{defina}) of the operator ${\cal Q}$. In spite
of some formal merits of such a step \cite{Batal} we are persuaded
that the related ``anisotropy" of both the operators ${\cal Q}$
and $\Theta_{(particular)}$ could hardly find a natural physical
foundation.

\subsection{The crossings and degeneracies of levels}

Although the Hamiltonian $H$ and ``spin" $\Omega$ (and
wavefunctions) of our model depend on three parameters, its energy
spectrum itself feels merely the influence of $Z$ and of the
product ${XY}$. In the light of eq.~(\ref{critin}) the two sources
of non-Hermiticity are the ``internal" strength $Z$ and the
``coupling" strength $\sqrt{XY}$. Still, the distinction between
$X$ and $Y$ is nontrivial. At $X=Y\neq 0$ giving a ``symmetrized"
coupling of channels, the operator of symmetry $\Omega$ becomes,
incidentally, Hermitian.

A strongly asymmetric decoupling of our model may be achieved by
the two alternative limiting transitions, viz., $Y \neq 0, \,X \to
0$ and $X\neq 0,\,Y \to 0$. Each of them suppresses just one of
the channels [cf. (\ref{bcmod})]. In both these limits the
symmetry $\Omega$ ceases to exist. One gets $\omega \to \infty$ or
$1/\omega \to \infty$ and our present method of solution becomes
inapplicable.

A much more interesting limiting transition $Z \to 0$ (from both
sides, i.e., $Z \to 0^+$ and  $Z \to 0^-$) converts our model into
a coupled set of two {\em Hermitian} square wells. In this limit
the violation of the Hermiticity  of the whole system is merely
caused by the channel-coupling terms. The energies $E = s^2 - t^2$
degenerate with respect to the spin since $t_\sigma(s) =
\sigma\,|t_\sigma(s)|$ so that, in fact, the neighboring $\sigma =
\pm 1$ levels cross at $Z=0$. No point of the crossing is
``exceptional" since the corresponding wavefunctions remain
linearly independent. Their Wronskian ${\cal W}$ does not vanish
and both our observables $H$ and $\Omega$ remain diagonalizable.
In contrast to some other solvable examples (say, to the harmonic
oscillator of ref.~\cite{ptho}), no Jordan-block structures emerge
in $H$.

For the sufficiently small $Z_{ef\!f}$, all the similar
observations may be made quantitative. Taking the ground-state
$n=0$ and setting $Z_{ef\!f} ={\cal O}( \lambda)$ while $Z ={\cal
O}( \lambda^2)$ for definiteness, we deduce that $t ={\cal O}(
\lambda)$ [cf. eq.~(\ref{rooty})]. Next we convert
eq.~(\ref{secular}) with $s=\pi/2+\varepsilon$ and a small
$\varepsilon=\varepsilon_0$ [cf. eq.~(\ref{nema}) in Appendix A]
in the leading-order estimate of $\varepsilon=4t^2/\pi+\ldots$.
All this transforms the definition of the energy into the
following approximate formula
 \ben
  E_0 = s_0^2 - t_0^2 =   \frac{\pi^2}{4}+ \frac{3XY}{\pi^2}-
  \sigma\,Z\,\frac{6\sqrt{XY}}{\pi^2}
  +{\cal O}\left ( \lambda^4 \right ).
  \een
As long as $t_\sigma(s) = \sigma\,|t_\sigma(s)|$, the ground state
has the ``spin" $\sigma=+1$ at $Z> 0$ and $\sigma=-1$ at $Z < 0$
while it becomes doubly degenerate at $Z = 0$. At this point the
Wronskian easily evaluates to a nonvanishing constant,
$
 {\cal W} =
 AC\kappa^* \sin 2\kappa^*\,
$. Hence, the two lowest states remain linearly independent at
$Z=0$.

\section{Summary}

The appeal of virtually all the PTSQM constructions may be seen in
a universality of their transition from a simple though indefinite
pseudometric $\theta$ to the correct physical and dynamically
determined positive-definite metric $\Theta$. The procedure is
counterintuitive and a number of open questions emerges. We
designed our present less trivial coupled-channel example to
clarify some mathematical subtleties (like the necessary
conditions of the reality of the spectrum in non-Hermitian
models), a deeper understanding of which requires, typically, a
nontrivial application of the Krein-space theory \cite{Langer}.

We believe that the explanation of many interrelated subtleties of
the PTSQM recipe may be facilitated via square-well models which
offer one of the most economical combinations of a transparent
dynamical picture with an exact solvability of the underlying
equations based on the usual matching technique. Our specific
present example illustrates, first of all, a phenomenologically
important situation where the dynamics is controlled by {\em more}
observable quantities.

An unexpected merit of our model has been found in a quick
convergence of the auxiliary perturbation expansions of its
energy-level parameters $s=s_n$ (such that $E=s^2-const/s^2$) in
the weak-coupling regime (i.e., for small strengths of the
non-Hermiticity $|Z|$ and $|XY|$) and/or in the quasi-classical
regime (i.e., at the higher excitations with $n \gg 1$). Another,
highly welcome byproduct of the square-well solvability emerged as
a non-Hermitian spin-type symmetry $\Omega$ of $H$. It enabled us
to reduce our nontrivial (viz., coupled-channel) Schr\"{o}dinger
equation to its much more easily tractable ``model-space"
reduction. In parallel, the existence of the symmetry enabled us
to analyze a level-degeneracy and level-crossing phenomena in a
neat, non-numerical manner.

The elementary algebraic structure of our model facilitated a
clarification of one of the most puzzling PTSQM requirements of
keeping {\em all} the observables $\Theta-$quasi-Hermitian and, at
the same time, $\theta-$pseudo-Hermitian in the Hilbert space
${\cal H}$. The coupled-channel (i.e., partitioned) structure of
the model enabled us to clarify the mechanism of this
correspondence anew. In particular, we showed that the
quasi-parity-based factorization $\Theta = \theta {\cal Q}$ as
introduced in ref.~\cite{psunit} appears mathematically more
natural than the alternative charge-based factorization $\Theta =
{\cal C}P$ of ref.~\cite{BBJ}, with $P=\theta$ in our present
notation. Indeed, while the quasi-parity is a symmetry of the
Hamiltonian itself (we have $[H,{\cal Q}]=0$), the formally
equivalent introduction of the charge ${\cal C}$ in~\cite{BBJ}
implies that $[H^\dagger,{\cal C}]=0$. This means that the charge
is merely a symmetry of an operator $H^\dagger \neq H$ defined as
a Hermitian-conjugate partner of the Hamiltonian.

Due to the existence of the second, $\theta-$pseudo-Hermitian and
$\Theta-$quasi-Hermitian spin-like observable $\Omega$ in our
model, another persuasive manifestation of a deep relevance of the
symmetries of $H$  has been revealed in the interrelations between
the full space Hilbert space ${\cal H}$ and its reduced,
single-channel subspace ${\cal H}_c$. {\it Pars pro toto}, the
quasi-parity-related factorization $\Theta=\theta{\cal Q}$ of the
metric in full space ${\cal H}$ has been proved accompanied by its
analogue (\ref{trubec}) using two relative quasi-parities ${\cal
R}(\sigma)$ defined within the single-channel subspace ${\cal
H}_c$.

The idea of the coupling of channels may turn attention to the
systems treated perturbatively in more dimensions \cite{Ema} as
well as to non-perturbative explanations of the observed
transitions between regular and chaotic classical and quantum
motion controlled by the partial differential equations (PDE)
\cite{Nana}. Via our example, some existing confirmations of the
internal consistency of the PTSQM theory may find their extension
to the coupled-channel scenario. On this basis, ``next" moves in
the PTSQM development may be predicted as aiming at the
non-separable PDE models \cite{Tater} where some aspects of our
model might inspire a more intensive exploration of the
level-degeneracy patterns in non-Hermitian context \cite{Tater2}
etc.

\section*{Acknowledgement}

Work supported by the grant Nr. A 1048302 of GA AS CR.


\newpage

\section*{Appendices}

\subsection*{A. Perturbation series for the energies}

PTSQM models usually require some sufficiently efficient numerical
description which is, typically, perturbative \cite{Bilape}. It
may mediate an alternative or quicker insight even in the solvable
models. {\it Vice versa}, the exact solvability of our present
model offers an explicit verification of the approximative
approaches.

The predominantly trigonometric oscillatory character of the
functions entering our secular eq.~(\ref{secular}) enables us to
locate and count all its physical roots,
 \be
 s=s_n=\frac{(n+1)\pi}{2}+(-1)^n\frac{Q_n}{2}
 , \ \ \ \ \ n = 0, 1, \ldots\
 \label{nema}
 \ee
where the new parameter $Q_n$ remains small in the weak-coupling
regime (i.e., for all the sufficiently small $X$, $Y$ and $Z$) as
well as at all the sufficiently large $n$. This enables us to
abbreviate
 \ben
  \frac{1}{(n+1)\pi}=\varrho
 \equiv \frac{1}{L},
 \ \
  \ \ \ \
  \frac{2\,Z_{ef\!f}(\sigma)}{L}=\alpha,
 \ \
  \ \ \ \
   \frac{2\,Z_{ef\!f}(\sigma)}{L^2}=\beta= \alpha\varrho
  \een
and to re-write our secular eq.~(\ref{secular}) in terms of these
new ``small" parameters and a sign factor $\tau = (-1)^{n}$,
 \be
  Q = {\rm arcsin}
  \left (2 t\, \frac{\varrho}{1+\tau\,Q\,\varrho}\,\sinh 2t\right ) , \ \ \ \ \ \
 2t = \frac{\alpha}{1+\tau\,Q\,\varrho}
 \,.
 \label{secularbez}
 \ee
As long as
 \ben
 {\rm arcsin}(x)=x+{\frac {1}{6}}{x}^{3}+
 {\frac {3}{40}}{x}^{5}+{\frac {5}{112}}{x}^{7 }
 +\ldots
 \een
we may iterate eq.~(\ref{secularbez}) and observe that the
$\alpha-$ and $\beta-$dependence of $Q=Q(\alpha,\beta)$ must
acquire the following general asymptotic-series form
 \be
 Q=Q(\alpha,\beta)=
 \alpha \beta\,
 \Sigma(\alpha,\beta), \ \ \ \ \ \ \
 \Sigma(\alpha,\beta)=
 \sum_{k,\ell=0}^{\infty}\,
 \alpha^{2k}\beta^{\,2\ell}\,c_{k\,\ell}\,
 \label{salve}
 \ee
where $c_{0\,0}=1$. In practice, this series should be truncated
in $\alpha$ as well as in $\beta$ or, equivalently, in $\varrho$.
Nevertheless, as long as the size of $\beta=\varrho\alpha$ is
dominated by $\alpha$, it is sufficient to analyse this series as
a power series expansion in the single ``small" parameter
$\alpha$. It is also worth noting that at any fixed power of
$\alpha$ there is always just a finite number of the related
powers of $\varrho$.

For illustration, let us set
 \ben
 \Sigma(\alpha,\beta)=
1+{\it c_{10}}\,{{\it\alpha}}^{2}+{\it c_{0\,1}}\,{{\it
\beta}}^{2}+{ \it c_{20}}\,{{\it\alpha}}^{4}+{\it
c_{11}}\,{{\it\alpha}}^{2}{{\it \beta}}^{2}+{\it c_{0\,2}}\,{{\it
\beta}}^{4} +{\cal O}(\alpha^6)
 \een
and insert this formula in the re-arranged eq.~(\ref{secularbez}),
 \be
 [ 1+\tau\, \beta^2
 \Sigma(\alpha,\beta)]\,
 {\rm arcsinh}\left \{
 [ 1+\tau\, \beta^2
 \Sigma(\alpha,\beta)]^2\,\frac{1}{\beta}\,\sin [\alpha \beta\,
 \Sigma(\alpha,\beta)]\right \} = \alpha
 \,.
 \label{secularse}
 \ee
As long as we can employ the regular Taylor series
 \ben
 \frac{1}{\beta}\,\sin [\alpha \beta\,
 \Sigma(\alpha,\beta)]=
{{\it \alpha}}+\left ({\it c_{10}}+{\it c_{01}}\,{{\it
\varrho}}^{2}\right ){{\it \alpha}}^{3}+\left [{\it c_{20}}+({\it
c_{11}}-1/6)\,{{\it \varrho}}^{2}+{\it c_{02}}\,{{\it
\varrho}}^{4}\right ]{{\it \alpha}}^{5}+{\cal O}\left ({{\it
\alpha}}^{7}\right )
 \een
the left-hand side of eq.~(\ref{secularse}) evaluates to a power
series in our small parameters. The tedious though straightforward
calculation converts the resulting equation into the infinite
series dominated by the leading-order identity
 \ben
 0=\left (-{\frac {1}{6}}+{\it c_{10}}+{\it c_{0\,1}}\,{{\it \varrho}}^{2}+3
\,\tau\,{{\it \varrho}}^{2}\right ){{\it \alpha}}^{3}+ \ldots\,.
 \een
It determines the first two coefficients,
 \ben
 c_{10}=\frac{1}{6}, \ \ \ \ \
 c_{0\,1}= -3\tau\,.
 \een
Their insertion simplifies the next-order $O\left ({{\it
\alpha}}^{5}\right )$ identity to the similar linear algebraic
relation which defines the next set of the coefficients in
$\Sigma(\alpha,\beta)$,
 \ben
 c_{20}=\frac{1}{120}, \ \ \ \ \
 c_{1\,1}=\frac{1-8\tau}{6}, \ \ \ \ \
 c_{0\,2}= 15\,.
 \een
In this manner one may continue the construction of the solution
(\ref{salve}) to an arbitrary order in $\alpha$.

In the original notation we may now write down the second-order
formula
 \ben
 Q_n=
 \frac{4\,Z_{ef\!f}^2}{(n+1)^3\pi^3}
  +
 \frac{8\,Z_{ef\!f}^4}{3\,(n+1)^5\pi^5}\,
 \left (
 {1}+\frac{18\,(-1)^{n+1} }{(n+1)^2\pi^2}
 \right )
 +{\cal O}\left (\frac{Z_{ef\!f}^6}{(n+1)^7}\right )
 \een
etc. The convergence in $1/(n+1)$ is amazingly rapid and the role
and weight of the non-Hermiticity decreases very quickly with the
growth of the excitation $n$.

In the single-channel limit where  $X=Y=0$ it has been observed
that the growth of $|Z|=|Z_{ef\!f}|$ makes some of the low-lying
energies move towards each other. With the growth of the absolute
value of $Z$ they first pair (in fact, $E_0$ and $E_1$) merges and
complexifies beyond the critical value of $Z_{crit} \approx 4.48$
\cite{sqw,sqwge}. In the coupled-channel context we may repeat the
same mathematical analysis leading, {\it mutatis mutandis}, to the
conclusion that all the observable values of energies $E=E_n$ and
quasi-spins $\sigma$ remain real in the weakly non-Hermitian
regime defined by the pair of inequalities $|Z_{ef\!f}(\sigma)|
 <
Z_{crit}$ , $\sigma = \pm 1$. They may be compressed into single
condition
 \be
 \left | \sqrt{XY} \right | + |Z|
 <
  Z_{crit} \approx 4.48\,.
  \label{critin}
 \ee
In contrast to the single-channel case, the energy spectrum now
ceases to be real at $Z =\pm \left ( Z_{crit}- \sqrt{XY} \right
)$, i.e., along two distinct surfaces in the space of parameters.

\subsection*{B. Modified Dirac's notation}



Our pair of operators $H$ and $\Omega$ samples a {complete} set of
non-Hermitian commuting observables. These operators enter
Schr\"{o}dinger eq.~(\ref{SE}) with quasi-spin constraint
(\ref{spinarch}), i.e., in the Dirac's notation, the pair of
equations
 \be
 H\,|E,\sigma\rangle = E\,|E,\sigma\rangle, \ \ \ \ \ \
 \Omega\,|E,\sigma\rangle = \sigma\,|E,\sigma\rangle
 \label{u}
 \ee
As long as $H^\dagger \neq H$ and $\Omega^\dagger \neq \Omega$,
the eigenkets $|\cdot,\cdot\rangle$ in (\ref{u}) {\em differ} from
the simultaneous eigenvectors of $H^\dagger$ and $\Omega^\dagger$.
In the spirit of ref.~\cite{KGA} let us now adapt the Dirac's
notation to the non-Hermitian scenario and equip the latter
elements of our Hilbert space ${\cal H}$ by the double delimiter.
The apparently unmotivated complex conjugation of the energies and
spins in their implicit definition
 \ben
 H^\dagger\,|E,\sigma\rangle\!\rangle = E^*\,|E,\sigma\rangle\!\rangle
 , \ \ \ \ \ \
 \Omega^\dagger\,|E,\sigma\rangle\!\rangle =
  \sigma^*\,|E,\sigma\rangle\!\rangle
 \een
becomes explained after the Hermitian conjugation which reveals
the ``left action" essence of these equations,
 \be
 \langle\!\langle E,\sigma|\,H=E\,\langle\!\langle E,\sigma|
 , \ \ \ \ \ \
 \langle\!\langle E,\sigma|\,\Omega=\sigma\,\langle\!\langle
  E,\sigma|
 \,. \
 \label{v}
 \ee
For this reason we shall prefer the use of the ket-vector form of
the `right' eigenfunctions $|E,\sigma\rangle$ in combination with
the doubly delimited (or `brabra-vector') form $\langle\!\langle
E,\sigma|$ of their `left-eigenfunction' partners.

Although both the latter sequences of vectors are defined,
strictly speaking, in the two equivalent copies of {\em the same}
Hilbert space of states ${\cal H}$, our notation conventions will
allow us to shorten the discussion here and there. As we already
emphasized, our ``redundant" version of the common Dirac's
notation is transparent and proves more consistent in the
non-Hermitian setting. Moreover, in the physical regime where
$E=E^*$ and $\sigma=\sigma^*$ the two pairs of Schr\"{o}dinger
equations (\ref{u}) and (\ref{v}) imply the {\em biorthogonality}
relations for their solutions which is easily written down now,
 \ben
 \langle\!\langle E',\sigma'|E,\sigma\rangle(E'-E)=0, \ \ \ \ \ \
 \langle\!\langle E',\sigma'|E,\sigma\rangle(\sigma'-\sigma)=0.
 \een
In the general non-degenerate case these rules only admit the
non-vanishing overlaps at $E'=E$ {\em and} $\sigma'=\sigma$. {\it
Vice versa}, unless one of the self-overlaps vanishes
accidentally, it is easy to derive the formal {\em completeness}
relations
 \be
 I = \sum_{E\!,\,\sigma}\
 |E,\sigma\rangle\,\frac{1}
 {\langle\!\langle E,\sigma|E,\sigma\rangle}\
 \langle\!\langle E,\sigma|\,.
 \label{completeness}
 \ee
Their use enables us to treat our set of two sequences of states
$\langle\!\langle E,\sigma|$ and $|E,\sigma\rangle$ as a
biorthogonalized basis giving straightforward formal expansions of
any element $|\alpha\rangle \equiv I \cdot |\alpha\rangle \in
{\cal H}$ or $|\beta\rangle\!\rangle \equiv I^\dagger \cdot
|\beta\rangle\!\rangle \in {\cal H}$. Thus, one derives
 \be
 H = \sum_{E\!,\,\sigma}\
 |E,\sigma\rangle\,\frac{E}
 {\langle\!\langle E,\sigma|E,\sigma\rangle}
 \langle\!\langle E,\sigma|, \ \ \ \ \ \ \
 \Omega = \sum_{E\!,\,\sigma}\
 |E,\sigma\rangle\,\frac{\sigma}
 {\langle\!\langle E,\sigma|E,\sigma\rangle}
 \langle\!\langle E,\sigma|
 \label{spectral}
 \ee
as two samples of an extension of the usual {\em spectral
representation} to (arbitrary) operators emerging in the
non-Hermitian coupled-channel context. These formulae will be
needed in sect. \ref{5.1}.

\subsection*{C. Partitioning of ${\cal H}$ into two
subspaces ${\cal H}_c$}

In eq.~(\ref{capibara}) we may employ the partitioned notation,
 \be
 |E_n,\sigma\rangle
  =\left (
 \begin{array}{l}
 |\varphi_n\rangle \cdot \sqrt{Y} \\
 |\varphi_n\rangle \cdot \sigma\,\sqrt{X}
 \ea
 \right ), \ \ \ \ \  \ \ \sigma = \pm 1, \ \ \ \  n = 0, 1,
 \ldots\,
 \label{bcmod}
 \ee
where the subkets $|\varphi_n\rangle$ are $\sigma-$dependent
solutions of eq.~(\ref{SEd}). In a left-action alternative to this
formula let us put
 \be
 \langle\!\langle E_n,\sigma| =
 \left (
 \sigma\sqrt{X}\, \langle \!\langle \chi_n|, \ \sqrt{Y}\,
  \langle \!\langle \chi_n|
 \right ), \ \ \ \ \  \ \ \sigma = \pm 1, \ \ \ \  n = 0, 1,
 \ldots\,
 \label{leftpar}
 \ee
where the new subcomponents $ \langle \!\langle \chi_n|$ are
defined by a left-action version of eq.~(\ref{SEd}). More
precisely, the left eigenstates $ \langle \!\langle \chi_n|$ and
the right eigenstates $|\varphi_n\rangle$ correspond to {\em the
same} reduced and spin-dependent parity-pseudo-Hermitian
single-channel sub-Hamiltonian
 \ben
 H(\sigma)=
 -\frac{d^2}{dx^2}+ V_a+\sigma\omega W_{b}
 = {\cal P}\,H^\dagger(\sigma)\,{\cal P}
 \,
 \een
which acts in the single-channel Hilbert subspace ${\cal H}_c$ and
which is, in the language of ref.~\cite{BB}, {\em truly} ${\cal
PT}-$symmetric.

The partitioning clarifies the structure of the spectral
representations of the operators in our basis. All of them may be
derived from the elementary projectors
 \ben
 |E,\sigma\rangle\,
 \langle\!\langle E,\sigma|=
  \left (
 \begin{array}{cc}
 |\varphi_n\rangle
 \cdot\sigma
 \sqrt{XY}\cdot
  \langle\!\langle \chi_n|&|\varphi_n\rangle
 \cdot Y \cdot
  \langle\!\langle \chi_n|  \\
 |\varphi_n\rangle
 \cdot X \cdot
  \langle\!\langle \chi_n|&|\varphi_n\rangle
 \cdot\sigma \sqrt{XY} \cdot
  \langle\!\langle \chi_n|
 \ea
 \right )
  \een
entering, say, eq.~(\ref{spectral}). They may be understood as
acting in two copies of ${\cal H}_c$ in ${\cal H}={\cal H}_c
\bigoplus {\cal H}_c$. We may abbreviate $|\varphi_n\rangle \equiv
|n\rangle$ and $\langle \!\langle \chi_n| \equiv \langle \!\langle
n|$ and collect the reduced Schr\"{o}dinger equations,
 \ben
 H(\sigma)|n\rangle = E_n\,|n\rangle , \ \ \ \ \
  \langle\!\langle n|\,H(\sigma)
  = E_n\,
 \langle\!\langle n|, \ \ \ \ \  \ n = 0, 1, \ldots\,.
 \een
They are to be solved in a  single copy of ${\cal H}_c$ where
their comparison leads to an alternative RN convention
 \be
 |n\rangle\!\rangle ={\cal P}\,
 |n\rangle\,{\lambda}_{E_n\sigma}^{(optional)}
 \label{starlet}
 \ee
paralleling eq.~(\ref{star}). We must check that and how both
these normalizations remain mutually compatible. For this purpose
we start from the RN definition (\ref{star}) and add the
partitioning (\ref{bcmod}) or, alternatively, start from the
partitioning (\ref{leftpar}) and insert the definition
(\ref{starlet}) afterwards. In the former case we proceed via
eq.~(\ref{capibara}) and get
 \be
 \langle\!\langle E,\sigma|E,\sigma\rangle=
  2\sigma\,
  \left ({\varrho}_{E_n\sigma}^{(optional)}
  \right )^*\,
 \sqrt{XY}\,
 \langle n |{\cal P}|  n \rangle\,.
 \label{subarujika}
 \ee
In the latter case we have
 \ben
 \langle\!\langle E,\sigma|E,\sigma\rangle= 2\sigma\sqrt{XY}\,
 \langle\!\langle n | n \rangle=
  2\sigma\,
  \left ({\lambda}_{E_n\sigma}^{(optional)}
  \right )^*\,
 \sqrt{XY}\,
 \langle n |{\cal P}|  n \rangle\,.
 \een
A comparison of these two results reveals that
 \be
  {\lambda}_{E_n\sigma}^{(optional)}
     \equiv {\varrho}_{E_n\sigma}^{(optional)}.
 \label{coincide}
 \ee
Our two apparently independent RN constants must be chosen equal
to each other.

\subsection*{D. Quasi-parity in the subspaces  ${\cal H}_c$}

When we move to the single-channel subspace ${\cal H}_c$ we
encounter the two different bases $\{ \, |n\rangle \,\}$
distinguished by the ``external" parameter $\sigma$. We have to
fix { $\sigma=+ 1$ or $\sigma=- 1$} in $|n\rangle=|n_\sigma
\rangle $. This means that in a sub-space analogue of
eq.~(\ref{defina}) we have to define the {\em two} ``reduced
quasi-parities" as operators ${\cal R}(\sigma)$ in ${\cal H}_c$
with a manifest dependence on the spin,
 \ben
 {\cal R}(\sigma)\,|n\rangle =
 {\cal R}(\sigma)\,|n_\sigma \rangle = |n\rangle\,
 {\varrho}^{(special)}_{E_n \sigma}\,.
 \een
At a fixed value of the spin $\sigma$ we obtain a subspace
counterpart of eq.~(\ref{expression}),
 \be
 \langle\!\langle {n'}_\sigma|n_\sigma \rangle_{(special)}=
 \langle {n'}_\sigma|{\cal P R}(\sigma)|n_\sigma\rangle\,,\ \ \ \
 \ \ \ \  n,n'=0,1, \ldots\,.
 \label{lexpressio}
 \ee
Although the spin-dependent product ${\cal P R}(\sigma)$ plays
just a not too important role of a subspace metric, its formal
prolongation from ${\cal H}_c$ to the full space ${\cal H}$ is
feasible and may be performed as follows. Firstly, one verifies
that the action of the $\sigma-$dependent auxiliary operator
 \ben
 {\cal S}(\sigma)=
 \left (
 \begin{array}{cc}
 0&\sigma \omega^{-1}\,{\cal R}(\sigma)\\
 \sigma\omega\,{\cal R}(\sigma)&0
 \ea
 \right )=
 \sigma\,\Omega\,{\cal R}(\sigma)
 \een
obeys the fixed-spin relation
 \ben
 {\cal S}(\sigma)\,
 |E_n,\sigma\rangle=\sigma\,
 {\cal R}(\sigma)\,|n\rangle \,\Omega\,\left (
 \begin{array}{l}
 \sqrt{Y} \\
  \sigma\,\sqrt{X}
 \ea
 \right )
  =
 |E_n,\sigma\rangle\,
 {\varrho}^{(special)}_{E_n \sigma}
 \,.
 \een
A transition to the spin-independent formula will be then most
naturally mediated by an introduction of the two two-by-two-matrix
projectors $\Pi_\sigma=(I+\sigma\Omega)/2$,
 \ben
 \Pi_\sigma\,{\cal S}(\sigma)\,\Pi_\sigma
 |E_n,\sigma\rangle=
 |E_n,\sigma\rangle\,
 {\varrho}^{(special)}_{E_n \sigma},
 \ \ \ \ \ \ \ \sigma = \pm 1.
 \,
 \een
We may conclude that the spin-independent quasi-parity operator in
the full space ${\cal H}$ may be {\em defined} by the formula $
{\cal Q}=\sum_{\sigma=\pm 1}\, \Pi_\sigma\,{\cal
S}(\sigma)\,\Pi_\sigma$. In the partitioned notation we may
re-write this operator in the matrix form,
 \ben
 {\cal Q}=
 \frac{1}{2}
 \left (
 \begin{array}{cc}
 {\cal R}(+1)+{\cal R}(-1)&
\omega^{-1}\,\left [ {\cal R}(+1)-{\cal R}(-1) \right ]
 \\
 \omega\,\left [{\cal R}(+1)-{\cal R}(-1)\right ]
 &{\cal R}(+1)+{\cal R}(-1)
 \ea
 \right ).
 \een
A return to another representation in terms of the projectors
$\Pi_\sigma$ is now possible,
  \be
 {\cal Q}
 =\sum_{\sigma=\pm 1}\,{\cal R}(\sigma)\,\Pi_\sigma
 \,.
  \label{defaaa}
 \ee
We see here that the two operators ${\cal R}(\sigma)$ may be
perceived as representing ``reduced" quasi-parities in ${\cal
H}_c$. The new version of the factorization formula for the metric
is delivered in the same spirit,
 \ben
 \Theta_{(special)} =
 \frac{1}{2}
 \left (
 \begin{array}{cc}
 \omega\,\left [{\cal P R}(+1)-{\cal P R}(-1)\right ]
 &{\cal P R}(+1)+{\cal P R}(-1) \\
 {\cal P R}(+1)+{\cal P R}(-1)&
 \omega^{-1}\,\left [ {\cal P R}(+1)-{\cal P R}(-1) \right ]
 \ea
 \right ).
 \een
In the light of eq.~(\ref{defaaa}) this formula represents our
factorized metric $\Theta_{(special)} = \theta \,{\cal Q}$ as a
weighted sum of two factorized items equipped with the appropriate
spin projectors,
 \be
 \Theta_{(special)} =
 \frac{1}{2}
 \,\sum_{\sigma=\pm 1}\,
  \left (
 \begin{array}{cc}
 \sigma \omega
 &1 \\
 1 &
 \sigma \omega^{-1}
 \ea
 \right )\,{\cal P R}(\sigma)
 =\sum_{\sigma=\pm 1}\, \left (
 \begin{array}{cc}
 0
 &{\cal P R}(\sigma) \\
 {\cal P R}(\sigma) &
 0
 \ea
 \right )\,
 \Pi_\sigma
 \,.
 \label{trubec}
 \ee
This conclusion is compatible with formulae (\ref{subarujika}) and
(\ref{expression}).



\newpage

\begin{thebibliography}{00}

\bibitem{BB}
Bender C M and Boettcher B 1998 Phys. Rev. Lett. 80 4243;

Bender C M, Boettcher S and Meisinger P N 1999 J. Math. Phys. 40
2201

\bibitem{DDT}
Dorey P, Dunning C and Tateo R 2001  J. Phys. A: Math. Gen. 34
5679

\bibitem{Shin}
Alvarez G 1995 J. Phys. A: Math. Gen. 27 4589;

Shin K C 2001 J. Math. Phys. 42 2513;

Mostafazadeh A 2005 J. Phys. A: Math. Gen. 38 6557

\bibitem{sgezou}
Andrianov A A, Cannata F, Dedonder J-P and Ioffe M V 1999
Int. J. Mod. Phys. A 14 2675;

L\'evai G and Znojil M 2000
J. Phys. A: Math. Gen. 33 7165

\bibitem{QES}
Bender C M and Boettcher S 1998 J. Phys. A: Math. Gen. 31 L273;

Znojil M 2000
J. Phys. A: Math. Gen. 33 4203;

Bagchi B, Cannata F and Quesne C 2000
Phys. Lett. A 269  79

\bibitem{sqw}
Znojil M 2001 Phys. Lett. A. 285  7

\bibitem{Kurasov}
Hern\'{a}ndez E, J\'{a}uregui A and A Mondrag\'{o}n A 2000 J.
Phys. A: Math. Gen. 33
4507;

Albeverio S, Fei S-M and Kurasov P 2002 Lett. Math. Phys. 59 227;

Weigert S 2004 Czech. J. Phys. 54 1139;

Znojil M and Jakubsk\'{y} V 2005 J. Phys. A: Math. Gen. 38 5041;

Fei S-M 2005 Czech. J. Phys. 55 1085

\bibitem{reviews}
dedicated January and October issues: 2004 Czech. J. Phys. 54, pp.
1 - 156 and 1005 - 1148 and

dedicated September issue: 2005  Czech. J. Phys. 55, pp. 1045 -
1192

\bibitem{ptho}
Znojil M 1999 Phys. Lett. { A 259} 220

\bibitem{Heiss}
Dembowski C et al 2001
 Phys. Rev. Lett. 86 787;

Heiss W D and Harney H L 2001 Eur. Phys. J. D 17 149;

G\"{u}nther U and Stefani F 2005 Czech. J. Phys. 55 1099

\bibitem{frag}
Znojil M 2004
J. Math. Phys. 45  4418

\bibitem{Bijan}
Bagchi B, Mallik S and Quesne C 2002 Mod. Phys. Lett. A 17 1651;

\bibitem{Batal}
Mostafazadeh A and Batal  A 2004
J. Phys. A: Math. Gen. 37 11645

\bibitem{robust}
Znojil M 2005
J. Math. Phys. 46 062109

\bibitem{Bila}
Quesne C et al 2005
Czech. J. Phys. 55 1161

\bibitem{Demiralp}
Demiralp E 2005 Czech. J. Phys. 55 1081

\bibitem{BG}
Buslaev V and Grecchi V 1993 J. Phys. A: Math. Gen. 26 5541

\bibitem{BBJ}
Bender C M, Brody D C and  Jones H F 2002 Phys. Rev. Lett. 89
0270401;
%
%
%

Kleefeld F 2003 AIP conf. proc. 660 325;

Znojil M 2004 hep-th/0408081
in "Symmetry Methods in Physics", CD ROM proc. series,
Ed. C. Burdik, O. Navratil and S. Posta (Dubna: JINR)

\bibitem{Critique}
Mostafazadeh A 2003 LANL
%
%
%
quant-ph/0310164;



Bender C M, Brody D C and  Jones H F 2004 Phys. Rev. Lett. 92
 0119902 (erratum)

\bibitem{ali}
Mostafazadeh A 2002 J. Math. Phys. 43 205

\bibitem{KG}
Vanroose W, Van Leuven P, Arickx F and Broeckhove J 1997 J. Phys.
A: Math. Gen. 30
5543;

Mostafazadeh A 2003
Class. Quantum Grav. 20  155

\bibitem{KGA}
Znojil M 2004 J. Phys. A: Math. Gen. 37 9557

\bibitem{sqwge}
Znojil M and L\'{e}vai G 2001
Mod. Phys. Letters A 16  2273;

Jakubsk\'{y} V and Znojil M 2004 Czech. J. Phys. 54 1101;

Weigert S 2005  Czech. J. Phys. 55 1183

\bibitem{Langer}
Langer H and Tretter C 2004 Czech. J. Phys. 54  1113


\bibitem{ook}
Mostafazadeh A 2002 J. Math. Phys. 43  2814 and 3944

\bibitem{Geyer}
Scholtz F G, Geyer H B and Hahne F J W 1992 Ann. Phys. (NY) 213
 74

\bibitem{Geyeru}
Geyer H B, Scholtz F G and Snyman I 2004 Czech. J. Phys. 54 1069;

Kretschmer R and Szymanowski L 2004 Phys. Lett. A 325 112


\bibitem{psunit}
Znojil M 2001 LANL report
math-ph/0104012, reprinted 2004 Rendiconti del Circ. Mat. di
Palermo, Ser. II, Suppl. 72  211


\bibitem{ja}
Bagchi B and C. Quesne C 2000
         Phys. Lett. A 273
                285;

Bagchi B, Quesne C and Znojil M 2001 Mod. Phys. Lett. A 16 2047;

Sinha A and Roy P 2004 Czech. J. Phys. 54 129

\bibitem{Ema}
Caliceti E 2004 Czech. J. Phys. 54 29;

Caliceti E 2005 Czech. J. Phys. 55 1077

\bibitem{Nana}
Ahmed Z and Jain S R 2003 Phys. Rev. E 67 R045106;

Nanayakkara A and Abayaratne Ch 2003
Can. J. Phys. 81 835

\bibitem{Tater}
Znojil M and Tater M 2001
J. Phys. A: Math. Gen. 34 1793;

Bender C M, Dunne G V, Meisinger P N and Simsek M 2001 Phys. Lett.
A 281 311;

Basu-Mallick B, Bhattacharyya T, Kundu A and Mandal B P 2004
Czech. J. Phys. 54 5;

Jakubsk\'{y} V 2004 Czech. J. Phys. 54 67

\bibitem{Tater2}
Nanayakkara A 2005 Phys. Lett. A 334 144;

B\'{\i}la H, Tater M and Znojil M 2005 Phys. Lett. A., Comment,
submitted

\bibitem{Bilape}
Caliceti E, Graffi S and Maioli M 1980 Commun. Math. Phys. 75 51;

Fern\'{a}ndez F M, Guardiola R, Ros J and Znojil M 1998 J. Phys.
A: Math. Gen. 31 10105;

Bender C M 2004 Czech. J. Phys. 54 1027;

B\'{\i}la H  2004 Czech. J. Phys. 54 1049;

Jones H F 2004 Czech. J. Phys. 54 1107


\end{thebibliography}
\end{document}